\documentclass[twocolumn,nofootinbib,amsmath,amssymb,aps,prd,balancelastpage,superscriptaddress]{revtex4-1}

\usepackage{color}
\usepackage[dvipsnames]{xcolor}
\usepackage[active]{srcltx}
\usepackage{amsmath,amsfonts,amssymb,amsthm,amstext,amscd,eucal,srcltx}
\usepackage{epsfig,graphicx,bm}
\usepackage{epstopdf, epsf}
\usepackage{dcolumn}
\usepackage{hyperref}
\usepackage{tensor}
\usepackage{lipsum}
\usepackage{appendix}
\usepackage{tikz}
\usepackage{caption}
\usepackage{subcaption}

\newcommand{\be}{\begin{equation}}
\newcommand{\ee}{\end{equation}}

\newcommand{\bse}{\begin{subequations}}
\newcommand{\ese}{\end{subequations}}
\newcommand{\bea}{\begin{eqnarray}}
\newcommand{\eea}{\end{eqnarray}}
\newcommand{\ba}{\begin{array}}
\newcommand{\ea}{\end{array}}
\newcommand{\bc}{\begin{center}}
\newcommand{\ec}{\end{center}}

\newcommand{\Fc}{\mathcal{F}}

\begin{document}

\vspace*{3mm}

\title{Background-induced complex mass states of graviton: quantization and tensor power spectrum}

\author{Anna Tokareva}
\email{tokareva@ucas.ac.cn}
\affiliation{School of Fundamental Physics and Mathematical Sciences, Hangzhou Institute for Advanced Study, UCAS, Hangzhou 310024, China}
\affiliation{International Centre for Theoretical Physics Asia-Pacific, Beijing/Hangzhou, China}
\affiliation{Theoretical Physics, Blackett Laboratory, Imperial College London, SW7 2AZ London, U.K.}

\begin{abstract}
\noindent
We start from the assumption that the theory of gravity can be formulated in terms of 4-dimensional action, and there are only 2 graviton polarization states, as in general relativity. It can be a non-perturbative effective action discussed in the asymptotic safety program or the result of some other UV modification of the general relativity making it a complete theory. From these general grounds, we study the properties of the graviton two-point function on top of cosmological de-Sitter space. We find that the no-ghost requirement formulated for the flat background does not necessarily hold for the de-Sitter space where the graviton two-point function can have an infinite number of complex-conjugate poles. We show that under certain stability conditions, their appearance doesn't contradict any fundamental principle. We discuss their observational consequences for the tensor power spectrum from inflation and for the stochastic gravitational wave background.
\end{abstract}

\thanks{Preprint: Imperial/TP/2024/AAT/2}

\maketitle

\section{Introduction}

In the most general theory of gravity, we expect the presence of all possible higher derivative and higher curvature terms. We assume that the theory admits the description in terms of effective action reproducing the loop corrections \cite{Donoghue:1995cz,Burgess:2003jk} or even non-perturbative results obtained in the framework of Asymptotically Safe gravity \cite{Percacci:2017fkn,Reuter:2019byg, Knorr:2021iwv, Fehre:2021eob, Platania:2022gtt,Saueressig:2023tfy,Saueressig:2023irs}. The similar action is the starting point of Infinite Derivative gravity \cite{Kuzmin:1989sp,Tomboulis:1997gg,Biswas:2011ar,Koshelev:2017ebj} proposed to heal the ghost appearing in a renormalizable quadratic gravity \cite{Stelle:1976gc}. In all these cases, the action contributing to the flat space graviton propagator can be reduced to \cite{1606.01250,1602.08475,1711.08864}

\begin{equation}
\label{action}
\begin{split}
    S=\int d^4 x \sqrt{-g} \left(\frac{M_P^2}{2} \,R+  \frac{1}{2} R\,\Fc_R (\square) R+ \right. \\ 
    +  \left. \frac{1}{2} W_{\mu\nu\lambda\rho}\Fc_W (\square) W^{\mu\nu\lambda\rho}\right)
\end{split}
\end{equation}
on top of the flat spacetime background. The graviton propagator can be extracted from the second-order perturbation of the action \eqref{action} after the $3+1$ decomposition \cite{2211.02070}, 
\begin{equation}
    h_{ij}^{TT}\left(1+2\frac{\square}{M_P^2} \Fc_W(\square)\right)\square h_{ij}^{TT}
\end{equation}
At this point, the minimal requirement for the healthy theory is the absence of any extra pole in the graviton propagator, except for the standard massless spin-2 graviton. The presence of zeros of the propagator is also forbidden because they will cause unexpected IR singularities in higher point vertices, such as graviton scattering. This applies to the propagator as a function of momentum on the whole complex plane because the poles corresponding to complex mass states lead to classical instabilities and causality violations \cite{Platania:2022gtt,Draper:2020bop,Platania:2020knd}. Thus, the function $1+2\frac{z}{M_P^2} \Fc_W(z)$ should have no poles and zeros, though it can have a branch cut along the positive real axis if quantum corrections are included and the action \eqref{action} represents a quantum effective action for gravity \cite{Fehre:2021eob}.

If we expect only massless graviton to be in the spectrum of the theory \eqref{action} on top of flat space, there is an additional requirement that guarantees the absence of scalar degrees of freedom \cite{2211.02070},
\begin{equation}
    3\Fc_R(\square)+\Fc_W(\square)=0.
\end{equation}
Thus, there is only one independent function determining the non-perturbative graviton propagator around flat space.

What happens if we fix the action for gravity for the flat background and then move to the maximally symmetric spacetime? According to the results of \cite{1606.01250,1602.08475}, the action determining the graviton two-point function can be written in the same form for any theory of gravity obeying the general covariance. Thus,
\begin{equation}
    \label{actionDS}
    \begin{split}
    S=\int d^4 x \sqrt{-g} \left(\frac{M_P^2}{2} \,R+ \frac{1}{2}R\,\Fc^{(\bar{R})}_R (\square)R+\right.\\ 
    +  \frac{1}{2} \left. W_{\mu\nu\lambda\rho}\Fc^{(\bar{R})}_W (\square) W^{\mu\nu\lambda\rho} \right)
    \end{split}
\end{equation}
Recall that
\begin{equation}
    R_{\mu\nu}=\frac{\bar{R}}{4}g_{\mu\nu},~~R^{\rho}_{\mu\sigma\nu}=\frac{\bar{R}}{12}\left(\delta^{\rho}_{\sigma}g_{\mu\nu}-\delta^{\rho}_{\nu}g_{\mu\sigma}\right).
\end{equation}
In general, the form-factors $\Fc^{(\bar{R})}$ contain the dependence on the background curvature, and they are different from the functions in \eqref{action}. Does the consistency of the theory require them to have no extra degrees of freedom around any dS background? This adjustment is possible but once there is a deviation of the background from maximally symmetric, the equations of motion would still contain higher derivatives. Thus, in any case, we have to deal with non-local infinite derivative theory, to properly describe, for example, cosmological perturbations. As the simplest case, we study here the possibility that the effective action of gravity expanded around de-Sitter space has already an infinite number of complex mass states, which typically happens in infinite derivative theory expanded over a non-trivial background \cite{2006.06641}. We find that, unlike the case of the flat background, they are not pathological under certain conditions on their masses and de-Sitter curvature. We propose a Bunch-Davies vacuum definition and consistent quantization of these background-induced complex mass states. We find the conditions under which the instabilities caused by these complex mass poles are temporary and do not destroy the background. We also study their effect on the tensor power spectrum in slow-roll inflation and show that they could lead to potentially observable gravitational wave signals. 

The paper is organized as follows. In the next section, we discuss how an infinite number of complex mass states emerge in a curved spacetime background. In Section 3, we show that these states have a well-defined Bunch-Davies vacuum in de-Sitter space and they can be consistently quantized. We discuss also the role of the classical stability condition which prevents the fields from instabilities on a boundary of de-Sitter space. In Section 4 we discuss the implications of the complex mass states for inflation and show that these fields can have observational signatures in stochastic gravitational wave background.

\section{Emergence of complex-conjugate masses}

Given that infinite derivative action expanded over a non-trivial background typically develops a set of the infinite number of perturbations with complex-conjugate masses \cite{2006.06641,Frasca:2022gdz}, it is important to understand in which cases they are safe from developing huge instabilities. The other tricky question is related to their quantization and definition of the vacuum. In this section, we address these issues for the simplest case of de-Sitter space.

In the case of de-Sitter space, which is the most relevant for inflationary cosmology, the Lagrangian determining the two-point function of graviton can be written in the same form for any theory of gravity, however, the functions could be different, depending on the interactions of gravitons. If the action \eqref{action} originally defined for the flat space has no additional interaction terms, the same action on top of de-Sitter space has the form \cite{Biswas:2016egy},
\begin{widetext}
\begin{equation}
\label{propDS}
    {(h^{\mu\nu})}^{TT}\left(-\square_{dS}+\frac{\bar{R}}{6}\right)\left(1-\frac{2}{M_P^2}\left(-\square_{dS}+\frac{\bar{R}}{3}\right)\Fc_W\left(\square_{dS}+\frac{\bar{R}}{3}\right)\right) h_{\mu\nu}^{TT}
\end{equation}
\end{widetext}
Here $\square_{dS}=\nabla_{\mu}\nabla^{\mu}$ is the d'Alambertian operator in de-Sitter space.

Let us show here the origin of the difficulty of tuning this action, such that there are no extra degrees of freedom emerging on top of de-Sitter. In the framework of infinite derivative gravity \cite{Kuzmin:1989sp,Tomboulis:1997gg,Biswas:2011ar,Koshelev:2017ebj}, the tree-level propagator is fixed to be ghost-free in flat space which implies
\begin{equation}
    \Fc_W(\square)=\frac{e^{\sigma(\square)}-1}{2\square}.
\end{equation}
Here $\sigma(\square)$ is an entire function. Here we use the mathematical result that for the entire function, the only way to avoid having roots and poles is to be presented as an exponent of some other entire function. Substituting this to Eq. \eqref{propDS}, we find that the poles of the propagator can be found as solutions to the algebraic equation
\begin{equation}
   e^{\sigma(z+\bar{R}/3)} =\frac{2 \bar{R}}{\bar{R}-3 z},
\end{equation}
or
\begin{equation}
    \sigma(z+\bar{R}/3)=\log{\left(\frac{2 \bar{R}}{\bar{R}-3 z}\right)}+2\pi i N
\end{equation}
Thus, the equation will inevitably have an infinite number of complex-conjugate roots. Also, it can have a few additional real poles (solutions for $N=0$) which are certainly not a desired feature, as they will be ghosts. However, they can be avoided given the proper choice of the function. Unlike real poles, the complex-conjugate ones seem to be a generic property of non-perturbative graviton propagators. 

Adding extra interaction terms to the flat space gravity action \eqref{action} can make these poles disappear on top of any de-Sitter background \cite{Biswas:2016egy,Koshelev:2023elc}. However, these interactions would certainly affect the power-counting renormalizability properties of the gravity action. Although for the exact de-Sitter one can adjust the tree-level Lagrangian, such that it eliminates the non-desired degrees of freedom, it seems to be impossible that this holds already for the cosmological FRW metric when the scalar curvature changes with time. Needless to say, for more complicated cases of inhomogeneous backgrounds we do not expect any improvements.  

For the described reasons, it is likely that in most general non-perturbative action (especially for theories that are non-local in the UV) we have to deal with infinite number of background-induced degrees of freedom with complex-valued masses. These degrees of freedom certainly lead to instabilities when considered on top of flat space. However, in the curved backgrounds, they still can be classically stable, see the consideration in \cite{2006.06641} for the case of infinite-derivative scalar field. 

\section{Quantizing graviton states with complex masses: Bunch-Davies vacuum in de Sitter and parabola of stability}

The higher derivative action \eqref{actionDS} around de-Sitter space can be decomposed to a sum of an infinite number of the field actions leading to the second order equation of motion for each of them. Recall that the Lagrangian
\begin{equation}
    \label{exp}
    L=\phi F(\square) \phi ,~~F(\square)=\prod_{k} (\square-m^2_k)(\square-(m^2_k)^*)e^{\sigma(\square)}
\end{equation}
 is equivalent to 
 \begin{equation}
 \label{product}
     L=\sum_k F'(m^2_k)\phi_1 (\square-m^2_k)\phi_1+F'(m^2_k)^*\phi_2 (\square-(m^2_k)^*)\phi_2.
 \end{equation}
Here $\phi_1$ and $\phi_2$ are complex scalar fields (2 degrees of freedom each) with the condition $\phi_2=(\phi_1)^*$ which guarantees that the action is real, and kills the 2 degrees of freedom. Thus, the overall Lagrangian is real and contains 2 degrees of freedom for 2 complex-conjugate values of the mass. Note that this decomposition is possible only due to the symmetries of de-Sitter space. If there is a deviation from the exact de-Sitter, this expansion will not be possible anymore. 

The same decomposition can be done for the spin-2 field as well. We will work here with the use of ADM decomposition and concentrate on tensor sector \footnote {The corresponding vector and scalar sectors will have the same properties due to the symmetries of de-Sitter. We study the case of the scalar part in more detail (including its possible mixing with inflaton perturbations in \cite{inprep_scalar}.} Recall that the metric has a form
\begin{equation}
ds^2=a^2(-\partial \eta^2 +d x_i dx^i+ 2 h^{TT}_{ij}dx^i dx^j),
\end{equation}
where $h^{TT}_{ij}$ is a transverse and traceless metric perturbation. It can be decomposed into two degrees of freedom with the use of the normalized polarization tensors $e^{A}_{ij}$,
\begin{equation}
   h^{TT}_{ij}=\Sigma_{A} \,e^{(A)}_{ij} h^{A}.
\end{equation}
Here we assume that the fields with complex masses emerging on top of de-Sitter space do exist, and develop quantum fluctuations, similar to the real mass fields. Here we consider the modes in the tensor sector,
\begin{equation}
    S=\frac{1}{8 M_P^2}\int d \eta d^3 x a^2\left(h^A {\cal P}(\square_{dS}) h^A\right),
\end{equation}
where 
\begin{equation}
    {\cal P}(\square_{dS})=2\frac{\square_{dS}}{M_P}\Fc^{\bar{R}}_W(\square_{dS})+1.
\end{equation}
As ${\cal P}(\square_{dS})$ can be decomposed similarly to \eqref{product}, we obtain the same action for each polarization $h^A$, and each pair of the complex masses $m_i^2$ and $(m_i^*)^2$. We can capture the factors ${\cal P}'(m^2)$ into the normalization of the complex-valued fields\footnote{It is essential here that the lagrangian can be written in terms of the complex-valued fields. Real-valued fields could not be rescaled by the complex-valued constant which reflects the fact that the ghost sign of kinetic term cannot be consistently captured into the normalization. Thus, the system under consideration can be written in terms of fields that are not ghosts.}. After this normalization, we get
\begin{equation}
\begin{split}
  S_i= \frac{1}{8 M_P^2}\int d \eta d^3 x a^2 \left(h^A_1 (\square-m^2_i)h^A_1+\right.\\
  \left.+h^A_2 (\square-(m^2_i)^*)h^A_2\right).
  \end{split}
\end{equation}
where the modes $h^A_2= (h^A_1)^*$ represent complexified fields that are related by the operation of conjugation.

How to define the vacuum state of this system? The action of each field resembles one of the massive scalar fields in de-Sitter. If we define $\chi=(a h)/(2 M_P)$ (here the indices $1, ~2$ are omitted for simplicity) and go to Fourier space we get a standard equation
\begin{equation}
\label{chieq}
    \chi''+\left(k^2-\frac{a''}{a}+m_+^2 a^2\right)\chi=0.
\end{equation}
Here $m_+^2=m_1^2+ i m_2^2$, $m_-^2=m_1^2- i m_2^2$. Recall that $\chi'=\partial_{\eta}\chi$, $a=-1/(H\eta)$. The vacuum state should be defined in the infinite past corresponding to $\eta\rightarrow -\infty$. The key point here is that in the limit $\eta\rightarrow -\infty$ the mass term is subdominant, and the solution is the same as for the massless field in the flat spacetime. Thus, the Bunch-Davies vacuum definition for each mode is the same as for the massless scalar,

\begin{widetext}
\begin{equation}
\chi({m^2}_+)=\sum_{i=1,2}\int \frac{d^3 k}{(2\pi)^{3/2}\sqrt{2 k}}\left({u_i}^+(\eta, {m^2}_+)
{\hat{A}^{\dagger\,(i)}}_k e^{-i k x}+{u^-}_i(\eta, {m^2}_+){\hat{A}^{(i)}}_k e^{i k x}\right).
\end{equation}
\end{widetext}

Here $[\hat{A}_k,\hat{A}^{\dagger}_q]=\delta^3(k-q)$ are the creation and annihilation operators, $u^+(\eta,_i m^2_+)\rightarrow e^{i k \eta}$ in the limit of $\eta\rightarrow -\infty$, which fixes a solution $u_+$ to Eq. \eqref{chieq} as
\begin{equation}
    u_i^+(\eta, m^2_+)=\frac{1}{2}(i+1)\sqrt{\pi k \eta} e^{i \pi n_+/2}(J_{n_+}(k\eta)+i Y_{n_+}(k \eta)),
\end{equation}
where we defined
\begin{equation}
    n_{\pm}=\frac{1}{2}\sqrt{9-4\frac{m_{\pm}^2}{H^2}}.
\end{equation}
The mode function $u_i^-$ can be found as a complex conjugate, $u_i^-(\eta,m^2_+)=(u_i^+(\eta,m^2_+))^*$. 

As the two degrees of freedom $\chi(m^2_-)$ are not independent but connected to $\chi(m^2_+)$ by the requirement of real-valued action, we can find them expressed through the same creation and annihilation operators $\hat{A}^{(i)}$,
\begin{widetext}
\begin{equation}
    \begin{split}
    \chi(m^2_-)=\sum_{i=1,2}\int \frac{d^3 k}{(2\pi)^{3/2}\sqrt{2 k}} \left(u_i^-(\eta, m^2_-){\hat{A}^{\dagger\,(i)}}_k e^{-i k x}+u^+_i(\eta, m^2_-)\hat{A}_k e^{i k x}\right).
    \end{split}
\end{equation}
\end{widetext}
Here the mode functions can be also computed as
\begin{equation}
     u_i^-(\eta,m^2_-)=(u_i^+(\eta,m^2_+)|_{m^2_+\rightarrow m^2_-})^*
\end{equation}

Thus, the power spectrum of each of the two degrees of freedom is
\begin{equation}
    P_{\chi}^{(i)}(\eta)=\frac{k^2}{2\pi}\left|u_i^+(\eta,m^2_+)(u_i^+(\eta,m^2_+)|_{m^2_+\rightarrow m^2_-})^*\right|^2.
\end{equation}
Finally, the tensor power spectrum generated by a pair of complex-conjugate masses is
\begin{widetext}
\begin{equation}
\label{PT}
P_{T}=\frac{4 H^2 q^2\eta^2}{(2\pi)^2}\,(-\pi q\eta)\,\left|e^{\frac{i\pi}{2}n_+}\mathcal{H}^{(1)}_{n_+}\,(-q\eta+ i\varepsilon)+e^{-\frac{i\pi}{2}n_+}\mathcal{H}^{(2)}_{n_+}\,(-q\eta-i \varepsilon)\right|^2\, 
\end{equation}
\end{widetext}
Here $\mathcal{H}^{(1),(2)}$ are Hankel functions of the first and second type, $\varepsilon >0$, $\varepsilon\rightarrow 0$ is introduced in order to select the relevant branch of the function. The additional factor 4 comes from 2 polarizations and 2 degrees of freedom with equal contributions. 

The power spectrum is convergent in the limit $\eta\rightarrow 0$ (i.e. on the boundary of de-Sitter) only if the condition 
\begin{equation}
\label{parabolaEq}
    m_2^2<3 \sqrt{H m_1^2}. 
\end{equation}
is satisfied ($H$ is the Hubble parameter characterizing the de-Sitter space). It was also derived in \cite{2006.06641} as a condition for classical stability of the scalar complex mass excitations on de-Sitter. It means that on the complex plane of $m^2$, the states should lie inside the parabola containing the positive ray of the real axis, see Figure 1. 
\begin{figure}[htb]
\label{parabola}
    \begin{center}
  \includegraphics[scale=.65]{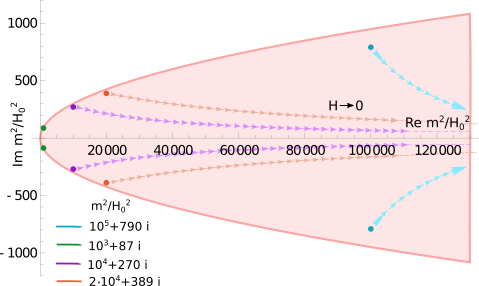}
  \end{center}
\caption{Illustration of the stability condition \eqref{parabolaEq} for complex mass states on inflationary de-Sitter. Arrows show schematically the expected behavior of the complex masses with decreasing spacetime curvature which guarantees stability of the theory of gravity on top of cosmological backgrounds. Recall that these states must disappear in the limit of the flat space. The selected points close to the stability constraint develop tensor modes slightly growing until the end of inflation. }
\end{figure} 

\section{Temporary instability implications for gravitational waves}

Summarising the result of the previous section, the background-induced complex mass states are safe from instability and can be consistently quantised\footnote{The relation of the complex mass states to the known classification of unitary representations in de-Sitter \cite{Joung:2007je} is tricky and deserves a special study. These states can be written as an infinite series of exceptional representations \cite{Bonifacio:2018zex}, however, their convergence properties are not obvious. We leave this question for future study.} if the values of the masses lie inside the parabola. Notice that when the curvature of de-Sitter space decreases, the parabola becomes closer to the real axis. In the limit of the flat space, complex-valued masses are not allowed.

It is expected that the values of the masses depend on the background curvature, although it might be difficult to extract this dependence because the full non-perturbative effective action for gravity is unknown. However, the stability condition applied for inflation stage in the early Universe puts a rough constraint on their curvature dependence. Namely, to have a safe transition to flat space all complex masses should asymptotically tend to the positive ray of the real axis and move to infinity when curvature goes to zero. This seem to be the only way to protect the gravity theory from developing instabilities in cosmological backgrounds. In what follows we will have in mind the described way of decoupling and disappearing of the complex mass states during the expansion of the Universe where curvature decreases with time.

Applying the obtained results to the cosmological (inflationary) de-Sitter one should remember that the Hubble scale slowly decreases closer to the end of inflation. This opens a possibility to get a few complex mass states moving closer to the stability parabola at the last stage of inflation. In general, this would lead to enhancement of the tensor modes for the short wavelength, and, in turn, to gravitational wave signals potentially observable by the future detectors \cite{1607.08697,1807.09495,1512.02076,astro-ph/0108011,Punturo:2010zz,1702.00786,2010.13157,1611.05560,1410.2334,2112.11465,2203.15668}. 

We illustrate this possibility using the example of the inflaton potential corresponding to the Starobinsky model \cite{Starobinsky:1980te},
\begin{equation}
    V(\phi)=V_0(1-e^{-2 \phi/(\sqrt{6}M_P)})^2.
\end{equation}
To a good approximation, in the slow roll regime, one can take a power spectrum obtained at the moment of horizon exit for each mode \cite{Mukhanov:1990me} using the slowly varying Hubble parameter which is also taken at this moment. It is convenient to parametrize the conformal momentum of each mode $q$ in terms of the number of e-foldings $N_q$ between the end of inflation and the horizon exit of $q$, $q \eta = e^{-N_q}$. For the Starobinsky potential, the Hubble parameter at $q$'s horizon exit is related to $N_q$ as
\begin{equation}
    H_q=H_0 (1-\frac{3}{2 N_q}).
\end{equation}
Thus the power spectrum can be obtained as
\begin{widetext}
\begin{equation}
\label{PT}
P_{T}=\frac{4 H_q^2 e^{-2 N_q}}{(2\pi)^2}\,(-\pi e^{- N_q})\,\left|e^{\frac{i\pi}{2}n_+(q)}\mathcal{H}^{(1)}_{n_+(q)}\,(-e^{-N_q}+ i\varepsilon)+e^{-\frac{i\pi}{2}n_+(q)}\mathcal{H}^{(2)}_{n_+(q)}\,(-e^{-N_q}-i \varepsilon)\right|^2\, 
\end{equation}
\end{widetext}

The expression in brackets has an approximate scaling
\begin{widetext}
\begin{equation}
    \left|e^{\frac{i\pi}{2}n_+}\mathcal{H}^{(1)}_{n_{+}}\,(-q\eta+ i\varepsilon)+e^{-\frac{i\pi}{2}n_+}\mathcal{H}^{(2)}_{n_{+}}\,(-q\eta-i \varepsilon)\right|^2\approx O(1) \frac{16 b^2}{\pi} (-q\eta)^{-2 a},
\end{equation}
\end{widetext}
where $b={\rm Im} \,n_+$ and $a={\rm Re}\, n_+ \approx 3/2$, $b\gg 1$.
Thus, near the stability parabola, the power spectrum increases proportionally to the real part of the mass, as $b^2\sim m_1/H_0$. Numerical plots confirm this asymptotic obtained from the series expansion for large complex index and small argument.

The gravitational wave energy density in the present Universe can be related to the tensor power spectrum as 
\begin{equation}
    \Omega_{GW}=\frac{1}{12}\Omega_{rad}\left(\frac{g^*_0}{g^*_{reh}}\right)^{1/3}\int \frac{d q}{q} P_T(q)
\end{equation}
where $\Omega_{rad}=2.47\cdot 10^{-5}$ is the energy fraction of radiation in the present Universe, $g^*_0$ and $g^*_{reh}$ are the effective numbers of relativistic degrees of freedom at reheating and now, correspondingly: $g^*_0=2 + 7/8\cdot(6\cdot4/11)$, $g^*_{reh}=106.75$.

\begin{figure}[htb]
\label{GW}
    \begin{center}
  \includegraphics[scale=.65]{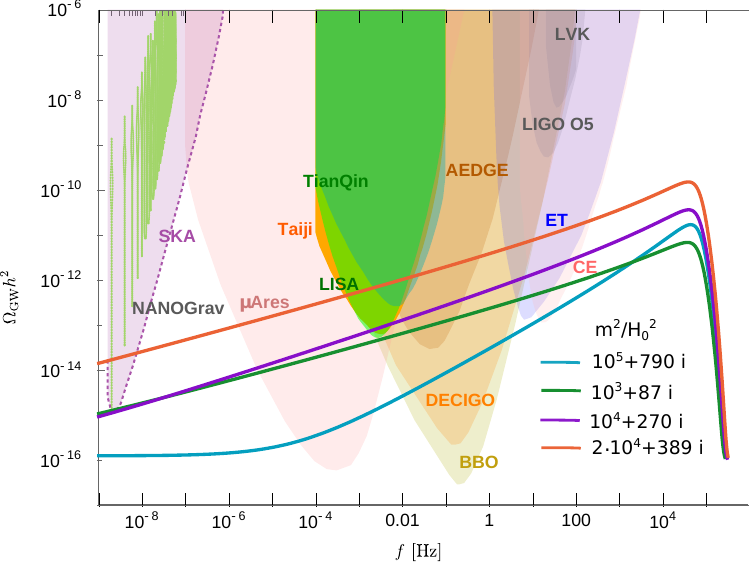}
  \end{center}
\caption{Gravitational wave spectrum generated by the temporary instability developing due to background induced complex mass states plotted over the projected sensitivities of the future GW detectors. Part of the projected limits were taken from the paper \cite{Marfatia:2023fvh} which includes a comprehensive summary of the current GW detection proposals.}
\end{figure} 

Our plots (see Figure 2) confirm the relevance of the effect of approaching the mass to the stability parabola for enhancement of the power spectrum at higher frequencies. Recall that in the range of conformal momenta measured and constrained by Planck \cite{Planck:2018jri} the spectrum cannot be significantly modified which means that at low frequencies ($N_q\approx 50$) all these modes should be well inside the stability region. In fact, this is a constraint on graviton interactions and non-perturbative action for gravity.

At high frequencies, the spectrum must get a cutoff related to the end of inflation and subsequent reheating.
In the case of immediate reheating the total tensor power spectrum can be found as a sum of the contributions from each complex mass state\footnote{This is not guaranteed if the dynamics of reheating is more complicated.}. It is relatively easy to get a sizable GW signal from the states with the real part of the mass around $10^2-10^3\, H_0^2$.

\section{Conclusions}
While the graviton two-point function must be fixed on top of the flat space, such that the dressed propagator has no additional (ghost) poles, it is not necessary to require the same tuning for any other background. Moreover, it seems to be not possible to achieve that for arbitrary backgrounds. We show that on top of de-Sitter space, an infinite number of complex mass states emerge as a typical phenomenon in generic quantum gravity action unless graviton interactions are specially tuned. We obtained the classical stability conditions for these background-induced complex mass states and proposed a quantization of these modes in a de-Sitter background. We found that the Bunch-Davies vacuum is well-defined at $\eta\rightarrow\infty$ for these states, as the mass term is subleading in the bulk limit of de-Sitter space. As the initial conditions must be complex-conjugate for the complex-conjugate masses, each pair of them corresponds to a single degree of freedom with its creation and annihilation operators. With this prescription, the classical stability at $\eta\rightarrow 0$ guarantees the well-defined and bounded power spectrum for each mode. Typically, if the imaginary part of the mass is much smaller than the real part, their contribution to the power spectrum of tensor modes is suppressed. 

We extend our discussion to the case of cosmological de-Sitter space where the Hubble rate is slowly decreasing. We show that if the complex mass is close to the boundary of the stability region it can modify the power spectrum at high values of momenta. This can lead to potentially observable consequences for the tensor power spectrum and lead to sizable gravitational wave signals. This way, in certain cases, the signatures of graviton couplings can be observed in the stochastic gravitational wave signal.

\begin{acknowledgments}
AT is indebted to A. Koshelev, A. Addazzi, K. Mkrtchyan, A. Platania for numerous valuable discussions. AT also thanks Shi Pi for sharing the plots advising on proposals for future GW detectors and their projected sensitivities. The work of AT was supported by by the National Natural Science Foundation of China (NSFC) under Grant No. 12347103 and STFC grant ST/T000791/1. The work of AT was also supported by a visitor grant from Albert Einstein Center for Fundamental Physics (AEC, Bern).
\end{acknowledgments}

\end{document}